# A Novel Helical Thin-Film Flow Diverter: Design, Fabrication, and Computational Assessment of Hemodynamic Performance

**Samuel Voß[1,2], Philipp Berg[1,3], Janneck Stahl[1,3], Daniel Behme[1,4], Gábor Janiga[1,2], Rodrigo Lima de Miranda[5], Eckhard Quandt[6], Prasanth Velvaluri[6,7a]**

[1]Research Campus STIMULATE, Magdeburg, Germany
[2]Department of Fluid Dynamics and Technical Flows, University of Magdeburg, Magdeburg, Germany
[3]Department of Medical Engineering, University of Magdeburg, Magdeburg, Germany
[4]University Clinic for Neuroradiology, University Hospital Magdeburg, Germany
[5]Acquandas GmbH, Kiel, Germany.
[6]Chair for Inorganic Functional Materials, Kiel University, Kiel, Germany
[7]Laboratory for Biomedical Microtechnology, University of Freiburg, Freiburg, Germany

[a]Corresponding author, email: prasanth.velvaluri@imtek.uni-freiburg.de

# Abstract

Flow diversion has become a key treatment modality for selected intracranial aneurysms, relying on the principle that a dense mesh of stent wires disrupts blood flow into the aneurysm sac, promoting thrombosis and vessel reconstruction. Despite its clinical success, a subset of patients experiences incomplete occlusion or complications. This study investigates innovative helical thin-film implants (HTFIs), aiming to evaluate their flow-diverting efficacy.

Highly resolved computational fluid dynamics simulations were performed on two representative patient-specific aneurysm models. Two HTFI design variants were tested at various configurations (two rolling angles and three deployment positions). A total of 28 unsteady hemodynamic simulations were performed, comparing six hemodynamically relevant parameters against the pre-interventional state and a conventional braided flow diverter.

The HTFIs induced significant changes in intra-aneurysmal flow. Both designs performed similarly overall, with the shorter configurations (smaller rolling angle) demonstrating superior efficacy. These achieved average hemodynamic reductions of 52.2 % and 58.4 %, outperforming the benchmark braided flow diverter device (47.4 %). Sensitivity to positioning was modest, with the best configuration showing an average variation of only 5.3 %, suggesting good robustness despite the helical design's heterogeneous porosity.

These findings indicate that HTFIs offer promising flow-diverting capabilities. With further refinement in design and hemodynamic optimization, these implants hold potential as a next-generation alternative for the endovascular treatment of intracranial aneurysms—especially in applications requiring compatibility with smaller delivery systems.

# 1. Introduction

Intracranial aneurysms are balloon like lesions on blood vessels in the brain. Such lesions can increase in size and rupture which can further cause death if not treated in a timely manner [1]. There are many different approaches available for the treatment of intracranial aneurysms such as: Surgical clipping [2], coiling [3], stent assisted coiling [4], and stand-alone flow diverter (FD) therapy [5]. Apart from surgical clipping the other treatments are carried out using minimal invasive therapy, where in the implant is delivered to the artery using a catheter. While the coiling and stent assisted coiling use intrasaccular implants to fill the aneurysm volume with soft platinum coils to reduce the blood flow, the stand-alone FDs operate on their own and reduce the flow into the aneurysm. Further, novel approaches include multiple intrasaccular implants that can be placed directly inside the aneurysm sac [6–8]. However, in cases of critical aneurysm geometries having wide necks, an intra-aneurysmal device can migrate after the placement, to address this a stent-assisted-coiling or braided FD-based treatment approach is preferred.

Braided FDs consist of multitude thin wires made of shape memory alloys that are woven into a cylindrical shape [9]. Their main advantages include self-expansion, flexibility, and stand-alone operation [10]. However, they have certain limitations such as the constant porosity of braided FDs cover side branches, which can lead to undesired stenosis or vessel occlusions. One of the reasons for this is increasing off-brand usage of small-diameter FD's for the treatment of distal aneurysms [11,12]. As the FDs were initially designed to treat aneurysms in proximal vasculatures [13] which have larger vessel diameters.

In a review of over 3,700 patients, Liu et al. found that using flow diverters (FDs) in distal arteries often led to branch occlusion or vessel narrowing, especially in the anterior cerebral artery, posterior communicating artery, and the M2 segment of the middle cerebral artery [11]. The authors advise doctors to be cautious when using FDs for distal aneurysms, particularly in small, distal vessels like M2, because these devices may cover branch vessels, disrupt flow dynamics, and in some cases provoke ischemia-like symptoms or additional complications. In such cases, one can benefit from miniaturized individualized implants that have potential to be scaled down to match distal geometries and vary their porosity (at aneurysm vs side branch) to reflect patients' geometry.

Furthermore, braided FDs also suffer from braid-related complications, a recent review with over 3500 patients on braided FDs mention device-related issues such as fish mouthing (3 %), device braid collapse (1 %), and device braid narrowing (7 %) [14]. The authors conclude that lack of standardization creates difficulties in identifying and associate such device failures to neurological outcomes. To enable safer treatments, diversifying from conventional braiding related fabrication technologies are key, and it should offer patient-specific design adaptations that reduce side branch occlusions and eliminate braid-related device complications.

One such novel alternative might be the helical thin-film implant (HTFI) [15] which is fabricated using microsystem technologies [16]. It provides additional advantages such as 1) higher design freedom (offering patient specific designs, [15,17]), and 2) the ability to integrate functional layers for increased radiopacity [18] and biosensing [19]. Previous thin film-based FDs have shown variable flow into the aneurysm and the side branch as compared to braided FDs [15]. While such variable flow was demonstrated in the previous study the investigated design could not be conformal along curved vessels of the phantoms. To address this and previous listed challenges of braided FD's this study presents a novel HTFI. In addition to having the capability of being patient-specific also allows for increased flexibly and can be miniaturized towards the use of aneurysm treatments in distal locations.

However, its increased freedom also poses a question to evaluate its capability of flow reduction and the dependence of deployment and aneurysm geometry on the porosity which reflects on flow reduction. Thus, computation fluid dynamics (CFD) can be used to get first-hand information about the flow variation of the novel HTFI that corresponds to a difference in placement and aneurysm geometry and compares it with standard braided FDs. Over the past years, CFD has become an established tool for investigating the initiation, growth, and rupture of intracranial aneurysms, offering detailed insights into the temporal and spatial characteristics of blood flow without posing any risk to patients [20,21]. Beyond fundamental research, CFD has also been widely used to study the hemodynamic effects of flow diverting implants, enabling the assessment of how different device configurations alter intra-aneurysmal flow patterns (e.g., [22–24]). By simulating these interactions in silico, CFD supports the design and development of novel endovascular devices, reducing reliance on costly and time-intensive prototype manufacturing and in vitro testing. A key objective in device development is achieving sufficient flow reduction within the aneurysm to promote blood residence time, intra-aneurysmal stasis, and the subsequent initiation of thrombosis and re-endothelialization of the arterial wall. Therefore, studies that correlate device-induced flow alterations with clinical outcomes are of particular interest. Based on this rationale, the present study follows the hypothesis that a novel implant must exhibit intra-aneurysmal flow reduction at least equivalent to, or ideally greater than, that of current state-of-the-art devices to be considered hemodynamically effective.

In this study, innovative implant designs based on the thin-film technology are investigated. CFD is used to assess the flow diverting performance of different thin-film-based stent designs and compare them to the pre-interventional state as wells as to a braided FD. In addition, the sensitivity of the new implant regarding optimal positioning is investigated since the design-inherent helical structure results in heterogeneous porosity. To our knowledge, this is the first study on the patient-specific hemodynamic quantification of helical thin-film-based FD efficacy for intracranial aneurysm treatment.

## 2. Methods

### 2.1 Data selection

Two representative intracranial aneurysms are selected for this retrospective study. They are part of a patient cohort that was treated with a flow-diverting DERIVO 2 embolization device (Acandis GmbH, Pforzheim, Germany) at the University Hospital Magdeburg. Aneurysms A and B are located at the side branches of Internal Carotid Artery, respectively. Both are side wall aneurysms in classical location and shape for flow diverter therapy, see Figure 1. Surface models are segmented and processed based on 3D digital subtraction angiography. For more information regarding image acquisition, segmentation procedure and general clinical information see Stahl et al. [25].

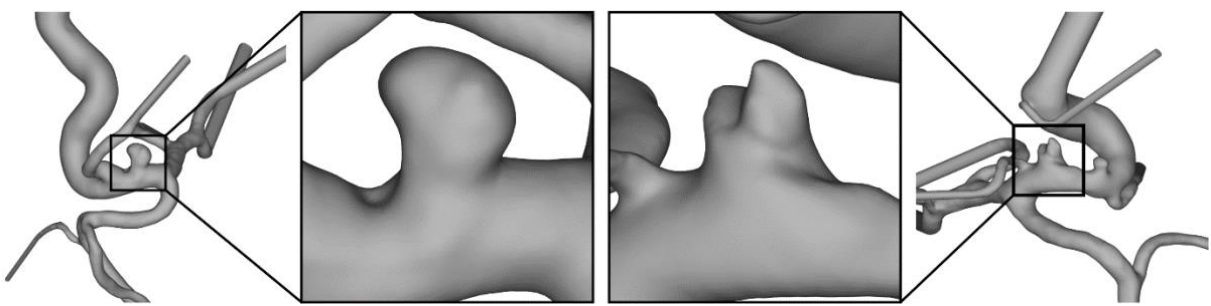

**Figure 1:** Aneurysm A and B and their respective location at the parent artery, side branches of the Internal Carotid Artery.

## 2.2 Design and fabrication

The novel helical implants are made from free standing nickel titanium (NiTi) alloys based thin-films and are fabricated using microsystem technologies such as UV lithography, magnetron sputtering, and wet chemical etching of sacrificial layers [16]. Complex geometries, with multiple thickness can be fabricated as demonstrated by Bechtold et al. [26]. Such complex structures are so called 2.5 dimensional geometries; they consist of thin-film NiTi of different thickness integrated monolithically. The design consists of two main parts: 1) The backbone and 2) the thin leaflets attached to the backbone (see **Figure 2**). The backbone is 42 µm thick and provides structural support to the implant. The thin leaflets are 7 µm thick and regulate the flow diversion capability of the implant. Its porosity can be modified as per the patient's requirement in the future.

Two designs are considered in this study: diamond and leaf structures. Initially, the fabricated devices are long strips, which are amorphous and flat. The devices are then wound circumferentially onto a stainless-steel cylinder at an angle to form a helix, the fixture holds the shape during thermal annealing. The two parameters that control the main dimensions of the device are the cylinder diameter and the rolling angle. Both determine the total length of the implant; and the pitch of the helix which reflects on the final device porosity. The devices along with the stainless-steel cylinders are annealed under vacuum at 550°C for 5 min. Post heat treatment, the device is both shape-set and crystallized that allows self-expansion.

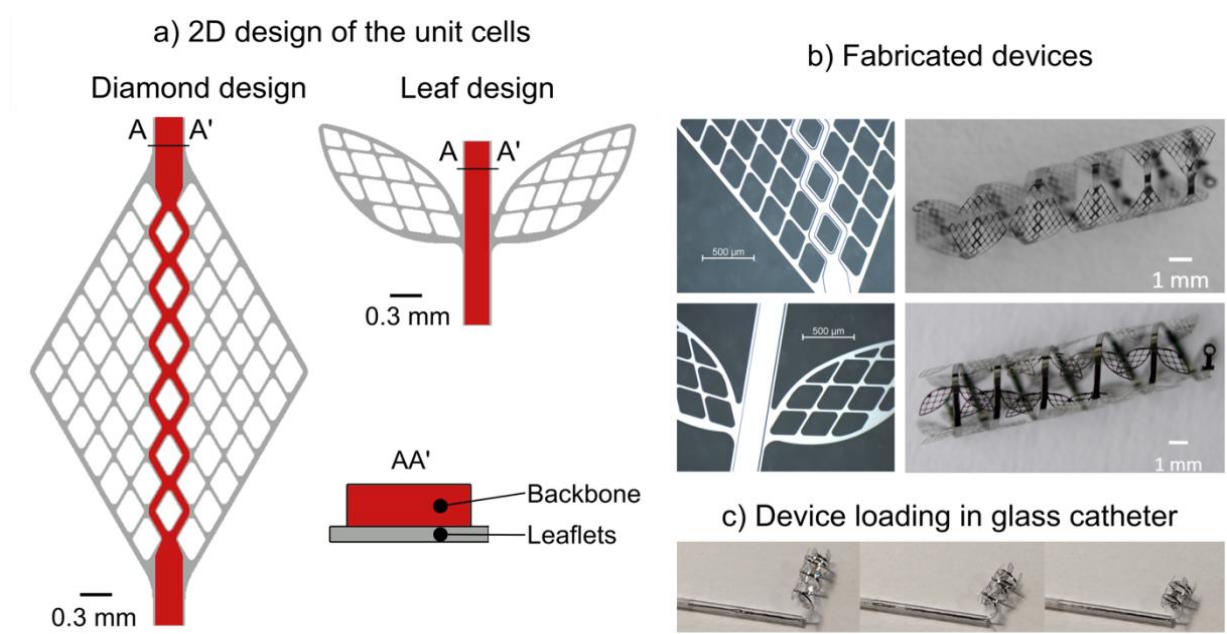

**Figure 2:** a) Design variants of the individual unit cells of the HTFI. Only the diamond design consists of porosity along the backbone. Inset shows the cross-section AA', the thick regions of the backbone (red) and the thin regions of the leaflets (grey). Both, the backbone and the leaflets, are made from NiTi. b) The devices are initially fabricated in a flat configuration. They are then rolled and heat-treated to set their final helical shape. c) During loading, the device is unrolled to fit into a delivery catheter of small diameter. Inside the catheter, the leaflets are folded. Upon deployment, the device is released, allowing the leaflets to unfold and the structure to recover its helical shape without compromising integrity.

## 2.3 Virtual stent variation and deployment

The braided FD serves as benchmark in this study. For aneurysm A and B a nominal diameter of 5 mm and lengths of 25 mm / 20 mm are employed, respectively [25]. For the HTFI, the rolling angle is the main design parameter as it leads to different stent lengths. The two rolling angles of 8° and 13° result in implant lengths of 9.1 mm and 13.3 mm for the diamond and 10.4 mm and 14.7 mm for the leaf design, further referend to as Length 1 and 2, see Figure 3. Short stents (compared to FD size) are used in this study to limit the computational effort. Longer stents are feasible to produce and compute but do not affect the ostium coverage. While braided FDs provide homogeneous porosity, the helical design of the HTFI lead to varying degrees of porosity and therefore aneurysm ostium coverage.

To assess the sensitivity of the device to positional variations, the implant orientation within the artery was systematically altered (Figure 3). Due to the helical architecture, the structural elements repeat periodically around the device circumference. In the diamond design, the main element recurs approximately every 150°, whereas in the leaf design, it recurs approximately every 60°. For each design, three orientations were investigated—0°, 50°, and 100° for the diamond, and 0°, 20°, and 40° for the leaf—covering a representative range of possible alignments. Both, braided FD and novel HTFIs, are virtually deployed using a fast virtual stenting approach [27]. Based on each aneurysm model's centerline the devices are deformed and further refined in an iterative process.

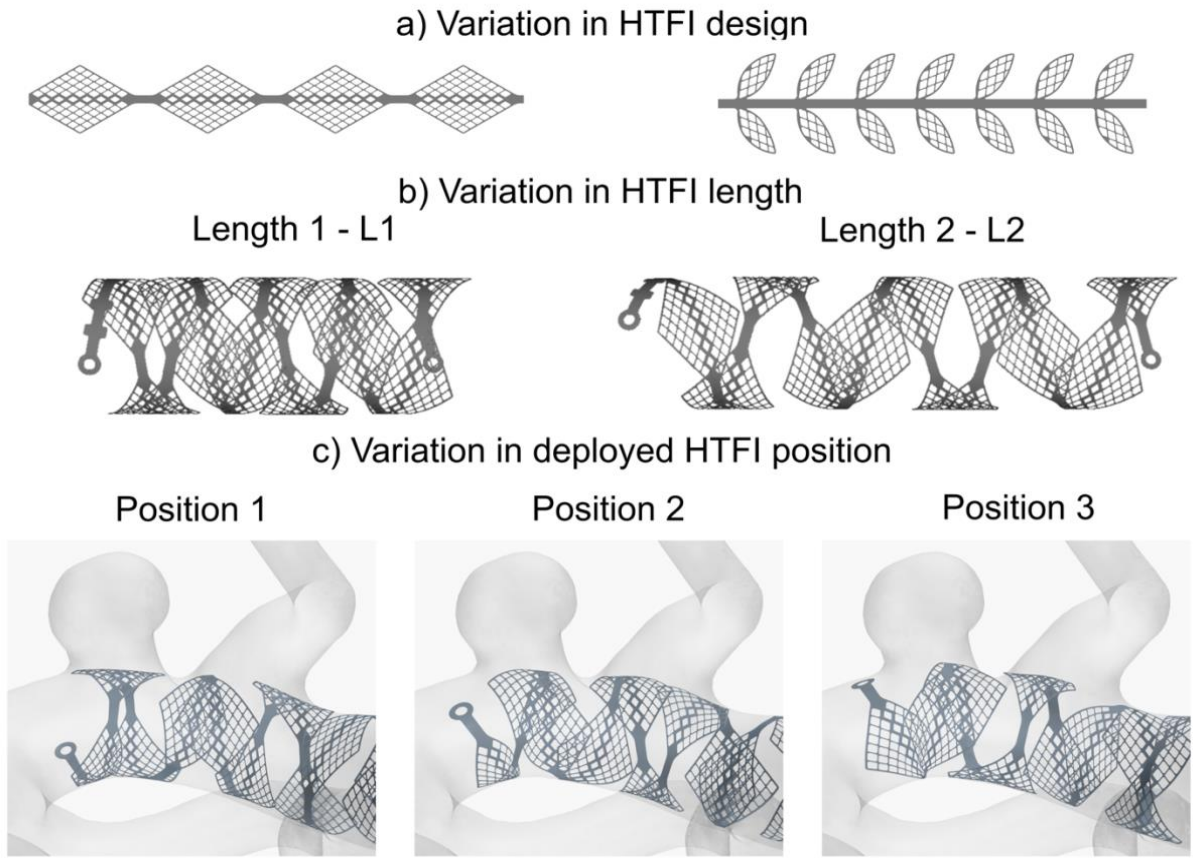

**Figure 3:** a) Two HTFI designs are compared (diamond vs leaf). b) For each design, two different rolling angles are used, determining the device length and final porosity. These are labelled as L1 and L2. c) Finally, for each design and each length there are three variations in terms of its position (Position 1-3) within the parent artery.

## 2.5 Hemodynamic simulation

All transient computational fluid dynamics simulations are performed inside the finite-volume solver STAR-CCM+ v17.06 (Siemens PLM Software Inc., Plano, TX, USA) following the simulation guidelines of [20,28]. For spatial discretization of the flow volume, polyhedral cells with a base size of 0.07 mm are used. In addition, five layers of prism cells at the wall boundaries ensure an accurate near-wall flow. Around stent wires and thin-film geometry, the discretization is further refined with a base size of 0.018 mm. This leads to a total cell count of 7.0 and 5.4 million for pre-interventional configurations, 25.4 and 13.6 million with FD and between 8.9 and 12.3 million for configurations with HTFI, respectively. A representative inflow curve [29] is applied at the inlet and scaled to the vessel diameter. Vessel walls are modeled as rigid with no-slip condition. For the pre-interventional state, an area-weighted outflow splitting is defined [30]. Then, the resulting time dependent pressure curves are used as outlet boundary conditions for the post-interventional state simulations. This approach allows an assessment of changes in the flow distribution caused by the implants. Blood is modeled as incompressible fluid ($\rho$ = 1056 kg/m$^3$) with laminar and non-Newtonian (Carreau-Yasuda model parameters taken from [31]) flow behavior. In total, 28 transient simulations are performed over two cardiac cycles. The first cycle serves initialization purposes, the second is used in the result evaluation.

## 2.6 Evaluation

First, simulation results are evaluated qualitatively based on flow field visualizations using velocity magnitude. Second, hemodynamic parameters are extracted to compare the HTFI with the pre-interventional configuration and the benchmark. For both, STAR-CCM+ software is used. Six hemodynamic parameters are analyzed:

- Mean velocity (Vmean, spatially averaged velocity inside the aneurysm),
- temporal velocity range (TVR, range between temporal minimum and maximum velocity),
- kinetic energy (KE, kinetic energy inside the aneurysm),
- neck inflow rate (NIR, volume flow rate into the aneurysm),
- inflow concentration index (ICI, degree of flow concentration when blood is entering the aneurysm) and
- mean wall shear stress (WSSmean, spatially averaged wall shear stress of the aneurysm wall).

MATLAB 2019b (The MathWorks Inc., Massachusetts, USA) is further used for calculating parameter reductions and creating bar plots.

## 3. Results

### 3.1 Qualitative results

Figure 4 presents qualitative hemodynamic results in terms of flow velocity at peak systole across all investigated configurations. The pre-interventional state and the case with a braided FD serve as reference and benchmark. In addition, the figure includes all configurations featuring the HTFI device, with variations in design (diamond and leaf), length (L1 and L2), and deployment position (P1-P3). Particular attention is given to the inflow region of the aneurysm (green arrow) and the adjacent side branch flow (purple arrow). Compared to the braided FD, configurations with HTFI show notable changes in intra-aneurysmal flow patterns and side branch perfusion. Among the HTFI designs, L2 exhibits a tendency toward higher flow velocities within both the aneurysm sac and the side branch, suggesting a less pronounced flow-diverting effect. In contrast, L1 demonstrates superior or comparable reduction in aneurysmal inflow, though it is associated with increased flow velocities in the side branch, potentially indicating reduced shielding of adjacent vessels.

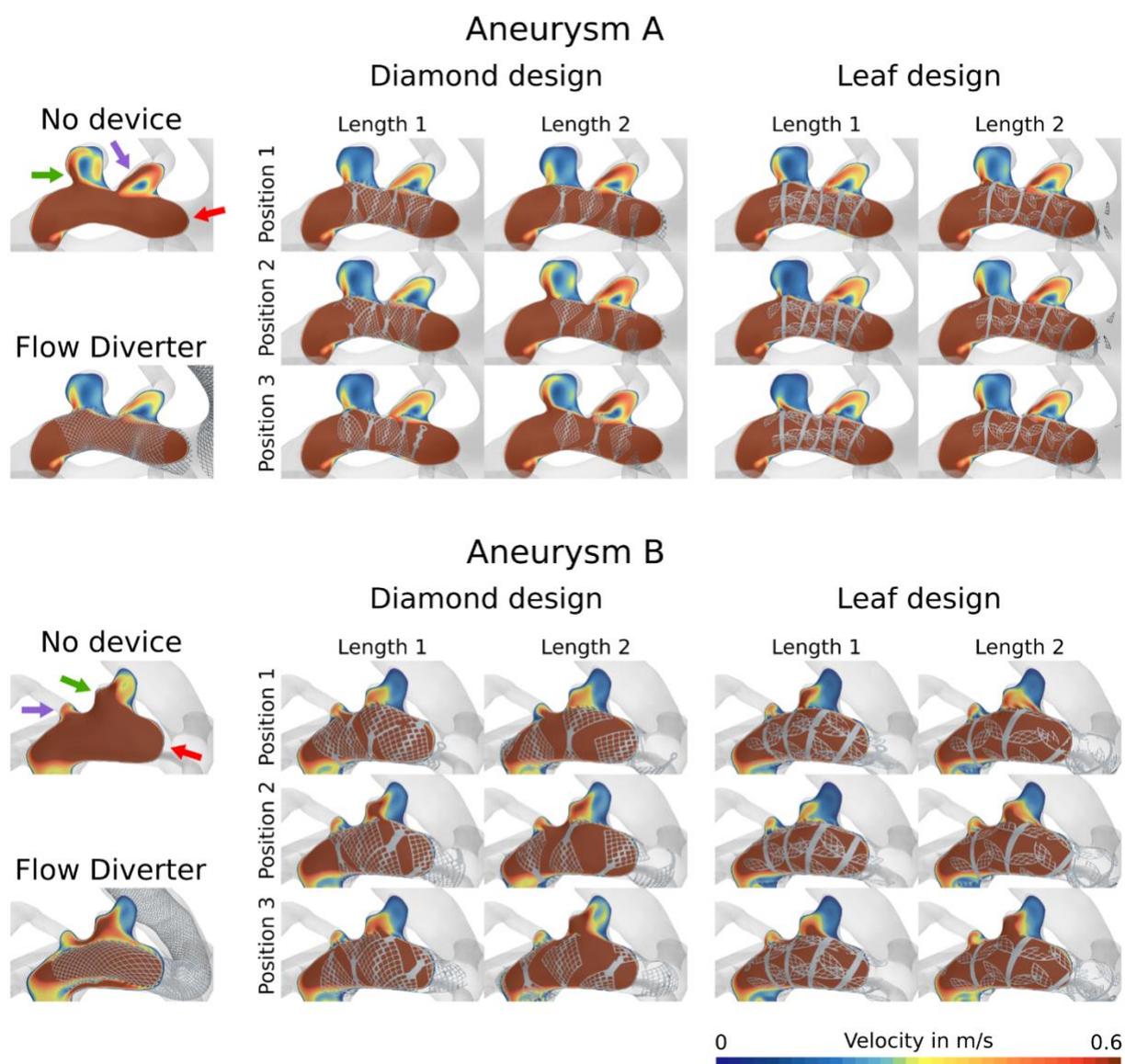

**Figure 4:** Comparison of pre-interventional state (no device) and braided FD with four HTFI designs (each in three different positions). Cross sections visualize the flow inside aneurysms A and B as well as their respective parent vessels by color-coded velocity magnitude. Arrows indicate the general flow direction (red), inflow region of the aneurysm (green) and side branch flow (purple).

## 3.2 Quantitative results

Table 1 summarizes the results for aneurysms A and B regarding six hemodynamic parameters that are used to characterize the intra-aneurysmal flow. Absolut parameters are given for the pre-interventional state. Further, the respective parameter reductions are listed for the braided FD and the HTFI configurations. Figure 5 shows bar plots of the parameter reduction by averaging the parameters of the three HTFI positions (P1-3), respectively; single values are indicated by circles. Both confirm the observations from Figure 4. In both aneurysms, the braided FD induces a similar reduction in key hemodynamic parameters, indicating a consistent and reproducible behavior across geometries. In contrast, the HTFI-based configurations show more variability, likely due to their less homogeneous porosity. This heterogeneity appears to affect the uniformity of flow reduction.

Across all metrics, the L1 variants consistently outperform the L2 counterparts. Specifically, for Vmean, KE, and WSSmean, both HTFI designs with L1 demonstrate performance comparable to the FD benchmark, while L2 designs fall short. When evaluating the TVR and NIR, all L1 configurations exceed the benchmark performance, whereas the L2 variants remain closer to benchmark levels, showing limited improvement. Regarding the ICI, the differences between configurations are less distinct, suggesting this parameter is less sensitive to design variation within the current range of HTFI configurations.

Furthermore, the influence of device positioning on hemodynamic parameters is assessed. To quantify this variation, the absolute deviation of each configuration from the mean value is calculated and averaged across the three positions. For the diamond design, the average variation is 5.3 % (±1.7 %) for Length 1 and 14.3 % (±4.6 %) for Length 2; for the leaf design, it is 5.9 % (±1.8) for Length 1 and 6.1 % (±3.6) for Length 2;

**Table 1:** Hemodynamic parameters for the pre-interventional state and the respective reduction for the FD and the HTFI in different design types (diamond vs leaf), Lengths (1 vs 2) and positions (P1-P3).

| | Parameter | Unit | Pre | FD | Diamond design of HTFI | | | | | | Leaf design of HTFI | | | | | |
| | | | | | Length 1 | | | Length 2 | | | Length 1 | | | Length 2 | | |
| | | | | | P1 | P2 | P3 | P1 | P2 | P3 | P1 | P2 | P3 | P1 | P2 | P3 |
|---|---|---|---|---|---|---|---|---|---|---|---|---|---|---|---|---|
| Aneurysm A | Vmean | m/s | 1.31E-01 | -55.0% | -60.3% | -47.2% | -59.8% | -61.3% | -28.2% | -28.3% | -48.6% | -63.4% | -62.6% | -41.9% | -38.4% | -43.7% |
| | TVR | m/s | 2.19E-01 | -31.5% | -36.3% | -35.0% | -41.2% | -32.4% | -12.5% | -24.4% | -41.9% | -50.4% | -47.9% | -27.9% | -27.3% | -35.4% |
| | KE | µJ | 2.54E-01 | -78.4% | -81.0% | -67.9% | -80.8% | -81.5% | -42.7% | -40.8% | -68.7% | -82.5% | -82.7% | -62.1% | -55.6% | -62.7% |
| | NIR | kg/s | 4.73E-04 | -37.5% | -43.2% | -36.9% | -44.7% | -44.6% | -26.9% | -32.1% | -46.6% | -53.6% | -50.7% | -30.6% | -33.4% | -38.3% |
| | ICI | - | 2.20E-01 | -48.0% | -53.0% | -41.0% | -55.0% | -58.1% | -36.0% | -30.3% | -47.1% | -53.5% | -58.3% | -33.1% | -32.6% | -38.2% |
| | WSSmean | Pa | 3.93E+00 | -66.2% | -66.6% | -54.0% | -66.4% | -69.5% | -25.9% | -19.2% | -51.7% | -66.1% | -68.6% | -43.9% | -32.6% | -40.7% |
| Aneurysm B | Vmean | m/s | 2.26E-01 | -42.8% | -59.0% | -41.4% | -52.4% | -57.7% | -46.2% | -18.2% | -52.3% | -69.8% | -54.4% | -59.9% | -44.4% | -32.6% |
| | TVR | m/s | 3.45E-01 | -26.0% | -47.6% | -32.4% | -40.2% | -45.6% | -31.3% | -12.5% | -44.3% | -56.9% | -43.8% | -43.5% | -31.9% | -26.6% |
| | KE | µJ | 3.47E-01 | -65.1% | -81.6% | -62.0% | -75.0% | -80.2% | -68.5% | -30.7% | -72.2% | -89.0% | -76.6% | -81.5% | -65.6% | -48.7% |
| | NIR | kg/s | 6.63E-04 | -36.5% | -49.3% | -43.9% | -44.0% | -47.7% | -43.8% | -21.0% | -51.1% | -64.0% | -50.2% | -51.6% | -40.1% | -34.1% |
| | ICI | - | 4.16E-01 | -31.9% | -48.2% | -33.1% | -43.3% | -48.0% | -38.7% | -1.5% | -45.6% | -63.7% | -44.9% | -53.9% | -39.0% | -26.7% |
| | WSSmean | Pa | 8.05E+00 | -50.5% | -58.4% | -40.2% | -58.4% | -61.0% | -45.7% | -13.1% | -49.9% | -72.2% | -57.4% | -62.6% | -44.5% | -24.5% |

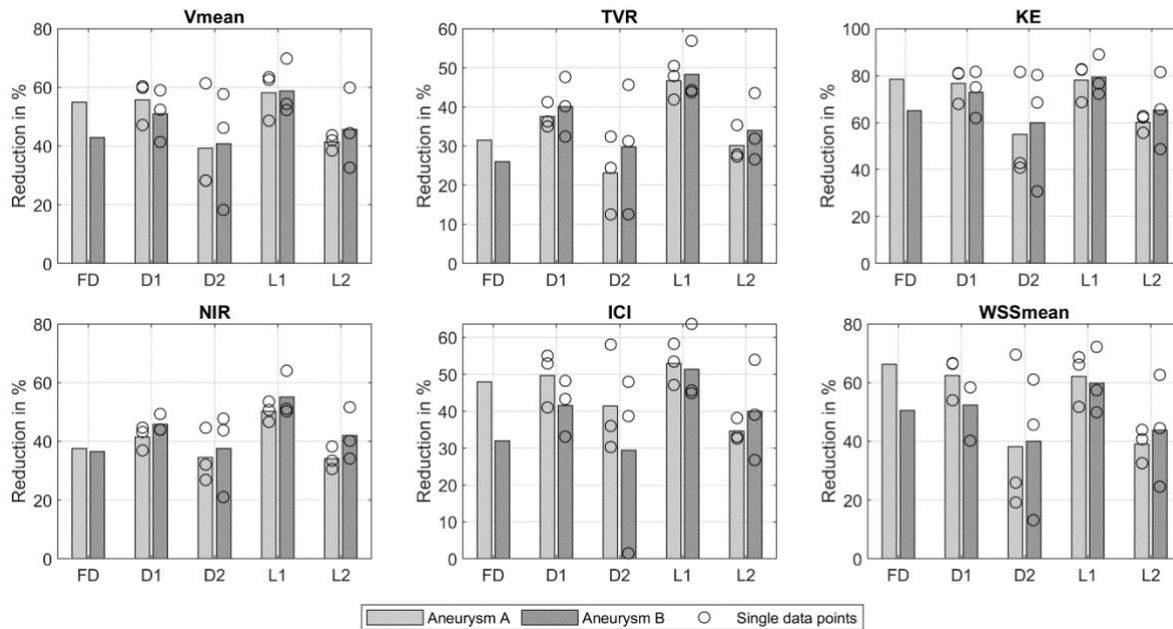

**Figure 5:** Bar plots showing the percental reduction compared to the pre-interventional state regarding six hemodynamic parameters for aneurysms A and B. Braided FDs are compared to the mean values four different HTFI configurations: Diamond design of Length 1 (D1) and 2 (D2), leaf design of Length 1 (L1) and 2 (L2). Each HTFI design is varied in position; circles indicate the individual results.

## 4. Discussion

The general flow modifications induced by the HTFI are encouraging and demonstrate the device's potential as an effective FD. Although the design types differ significantly in geometry, their influence on flow diversion is minor. Shorter and thus denser devices consistently perform better, which aligns with expectations. The novel helical structure shows a distinct sensitivity to positioning, unlike classical braided designs. For the better-performing short variants of the diamond and leaf designs, performance variability remains low at 5.3 % (±1.7 %) and 5.9 % (±1.8 %), respectively. These results are promising, especially given that deployment of braided FDs allows for even greater control over local density through compression [22,32].

However, a critical question remains: is the observed flow diversion sufficient to suggest a successful clinical outcome, specifically aneurysm occlusion? To address this, a classical braided FD is included in the comparison to serve as a clinically validated benchmark. This approach aligns with the hypothesis underpinning the present study: A novel implant must achieve intra-aneurysmal flow reduction at least equivalent to, and ideally exceeding, that of current state-of-the-art devices to be considered hemodynamically effective. The compact HTFI configurations (L1) show consistently equivalent or even superior performance relative to the classical device, particularly in terms of NIR and TRV. This suggests that specific HTFI designs may possess adequate hemodynamic efficacy to promote thrombus formation and eventual occlusion.

To further validate this assumption, results are compared to previous studies that explicitly differentiate between successful and unsuccessful treatment outcomes based on clinical follow-up and CFD simulations. These works provide a useful reference for estimating the hemodynamic thresholds required for successful aneurysm treatment. Mut et al. investigated the relation of intra-aneurysmal hemodynamic conditions after FD deployment and the occlusion time of a cohort primarily consisting of large aneurysms [33]. 23 aneurysms were included, 15 showed fast, the remaining a slow occlusion (patent or incomplete occlusion after 6 month). In the group of fast occlusion, significantly lower post treatment values occurred for Vmean and NIR. Average reduction from pre to post treatment state were 68 % and 87 % for the fast group and 28 % and 71 % for the slow group, respectively. In the present study, reductions in Vmean are comparable; however, the HTFI does not reach the same level of reduction in NIR observed in the fast-occlusion subgroup. In a subsequent study, the focus shifted to aneurysms treated with intrasaccular devices [34]. The investigation compared 18 completely occluded aneurysms to 18 incompletely occluded ones. The Vmean, NIR, and ICI were found to be reduced by an average of 81%, 48%, and 46%, respectively, in the completely occluded group. In contrast, the incompletely occluded group showed smaller reductions of 63 %, 10 %, and 8 % for the same parameters. The flow diverting effect in the intrasaccular devices was similar to those seen with the HTFI. Notably, the HTFI configurations in this study fall well within the average reductions of successfully occluded aneurysms for NIR and ICI. However, the Vmean remains higher than the successful group, which may partially reflect differences in implant mechanics or aneurysm morphology.

Stahl et al. compared pre- and post-interventional hemodynamics. FDs are used to virtually replicate the treatment of ten intracranial aneurysms [25]. They found average reductions of NIR (51 %), ICI (56 %), WSSmean (47 %) and KE (71 %) for aneurysms in the post-interventional state. These findings align closely with the values observed in the current study, further supporting the hypothesis that the HTFI achieves flow diversion within a clinically effective range.

Despite these promising comparisons, several factors complicate direct interpretation. Strong dependencies on boundary conditions [35], patient-specific size and shape variations [36], and the use of cohort-averaged flow reductions in prior studies limit the precision of cross-study comparisons. Nevertheless, the convergence of findings across different methodologies and aneurysm types provides substantial support for the potential clinical utility of the HTFI, especially when further refined through targeted design optimization.

Beyond the CFD findings a distinctive noteworthy advantage is the design of the proposed HTFI which allows it to be loaded into a catheter similar to a coil (see Figure 2). So, essentially it can be stored in a much smaller coil catheter but with the function of a FD. Its properties allow the device to be loaded into catheters which are slightly bigger than the width of the backbone (0.3 mm) for e.g., 1.7-1.8 Fr (not demonstrated here) beneficial for the treatment of distal aneurysms which are beyond the circle of Willis. Such aneurysms are challenging with braided FDs to treat mainly due to the smaller diameters and the distal navigation to such blood vessels [12,37,38]. Treatment of such aneurysms with braided devices has a complication rate of 10 % (in a study with over 168 patients) due to ischemic events and side branch occlusion [38]. In such cases, one can potentially benefit from HTFI as it can allow further miniaturization (as the fabrication is based on microsystem technology), patient-specific design (reducing side branch occlusion), and delivery with smaller catheters.

This study has several limitations that should be acknowledged. First, the limited number of representative aneurysm cases restricts the generalizability of the results. Additionally, the fast virtual stenting approach employed in this work is based on simplifications and does not account for the detailed mechanical behavior of the implants. Despite this, the method enables geometrically realistic deployment, as confirmed by prior validation studies [39,40]. Another limitation lies in the use of a generalized flow curve as the inlet boundary condition. While this approach simplifies the simulation setup, it does not capture the patient-specific variability in flow rates. However, since identical inlet conditions are used across all configurations, relative comparability between cases remains valid. Furthermore, the assumption of rigid vessel walls may neglect some biomechanical interactions; nevertheless, this is considered a reasonable simplification, as small vessel deformations have been shown to exert only limited influence on hemodynamic outcomes [41]. Finally, only two HTFI designs were investigated in this study, limiting the explored design variability.

Future research aims to address these limitations. Expanding the study cohort will enhance the applicability of findings to a wider range of anatomical and clinical scenarios. In vitro validation using patient-specific vascular phantoms could provide a more comprehensive assessment of the simulation accuracy and stent deployment realism. Lastly, incorporating parametric optimization techniques and local porosity variation may facilitate further improvements in device design.

## 5. Conclusions

This study demonstrates that the HTFI possesses promising flow diverting properties. Even in its preliminary design stage, configurations with smaller rolling angles and lower density perform at a benchmark level comparable to classical braided FDs, with a trend toward improved performance. However, the results also indicate increased variability, likely due to the inherent sensitivity of the helical geometry to rotation and positioning within the vessel. Given that the current HTFI design represents an early-stage prototype, these findings provide an encouraging outlook for its future development. With further design refinements and optimization, the HTFI may have the potential to surpass the performance of current state-of-the-art FDs. Assuming strong flow-diverting performance combined with unique advantages such as compatibility with small-sized catheters, the success of this novel device will ultimately depend on additional factors including mechanical properties, cost, navigability and patient safety. If these aspects prove favorable, the HTFI may offer a valuable alternative for the treatment of intracranial aneurysms.

## Data availability

The data that support the findings of this study are available from the corresponding author upon reasonable request.

## Acknowledgements

## Author contributions

[If required]

## Funding

This study was partly funded by the Federal Ministry of Research, Technology and Space in Germany within the Research Campus STIMULATE (grant number 13GW0473A and 13GW0674C) and the European Regional Development Fund (ZS/2023/12/182010).

## Competing interests

The authors declare no competing interests.

## Supplementary information